\documentclass[journal]{IEEEtran}
\usepackage{cite}
\usepackage{amsmath,amssymb,amsfonts}
\usepackage{booktabs}
\usepackage{graphicx}
\usepackage{textcomp}
\usepackage{url}
\usepackage{xcolor}
\usepackage[hidelinks]{hyperref}

\definecolor{orange}{rgb}{0.90,0.42,0.08}
\hypersetup{pdftitle={Selection-Aware Certification of Client-Constrained Fraud Rule Portfolios Under Synthetic Distribution Shift},pdfauthor={Rishabh Pande, Yu Zeng}}

\begin{document}
\title{Selection-Aware Certification of Client-Constrained Fraud Rule Portfolios Under Synthetic Distribution Shift}
\author{Rishabh Pande and Yu Zeng%
\thanks{Both authors are with Socure, United States (e-mail: rishabh.pande@socure.com; jason.zeng@socure.com). Rishabh Pande ORCID: 0009-0004-1901-7109. Corresponding author: Rishabh Pande.}}
\maketitle
\begin{abstract}
Selecting fraud-rule portfolios to satisfy a precision or recall floor can overfit the data used for selection. We study finite-family certification for bounded OR portfolios, comparing point-estimate selection, a fixed safety margin, and one-sided Clopper--Pearson lower bounds with Bonferroni adjustment. In a factorial follow-up with 6,400 datasets from two author-designed synthetic generators, full-validation certification reduced the held-out floor-violation rate by 29.1 percentage points on average versus full-validation point selection, while reducing deployment by 31.7 points. Compared with a fixed 0.05 margin, certification reduced violations by 22.2 points and deployment by 29.4 points; results varied across conditions. Using the full rather than half validation sample increased deployment by 17.1 points on average, without a uniform reduction in shifted-test violations. In a separate 400-dataset no-shift sweep, certification deployment rose from 61.3\% to 90.2\% as validation size increased from 1,500 to 15,000; no violations occurred among 2,887 deployed portfolios tested on 100,000-row holdouts. A temporal case study on the public synthetic Bank Account Fraud Base dataset used recall at a 5\% false-positive-rate cap; the selected candidate pool captured less than 1\% of test fraud, showing limited coverage in this setup. Together, these experiments characterize how finite-family certification changes the reliability--deployment tradeoff for bounded OR portfolios under specified synthetic conditions. The evidence covers the evaluated generators and one public synthetic BAF case study; it does not estimate performance on real transaction populations or robustness to arbitrary distribution shift. The authors revised and rewrote this abstract after receiving a draft from OpenAI Codex \cite{openai_codex}.
\end{abstract}

\section{Introduction}
Fraud detection systems trade off missed fraud against false alerts that impose investigation and customer-friction costs. A fraud team may specify a performance floor and a priority metric. We use \emph{precision-floor/recall-maximizing} for the policy that requires minimum precision and then maximizes recall, and \emph{recall-floor/precision-maximizing} for the reverse policy. These names identify the constrained metric directly. We assess both deployment frequency and the rate of floor violations among deployed portfolios.

Standalone rule metrics do not determine the value of a combined strategy. Two individually strong rules may cover many of the same transactions, while a rule with modest standalone performance can add unique fraud coverage. The compound precision and recall must be computed on the unique union of matched transactions. A held-out evaluation is also needed because a strategy that meets the chosen target during development may fail to do so on a later or shifted population.

Optimizing a finite-sample validation metric can also create selection bias: noise in the metric affects which portfolio is chosen, and the selected portfolio's score on the same data is not an unbiased estimate of future performance. This effect is established in model selection and cross-validation studies \cite{cawley2010selection,varma2006bias}. Our comparisons separate training-based ranking from certification and include a full-validation point selector to measure the consequences of reusing validation data for both ranking and selection.

Prior fraud-rule research studies automatic candidate generation, logical OR rule-set selection, and precision--recall tradeoffs. Statistical risk-control work also studies selection from a finite family using independent calibration data and multiplicity-aware testing. The closest framework, Learn-then-Test, casts risk control as multiple hypothesis testing and can certify sets of candidate procedures \cite{angelopoulos2025learntest}; post-selection confidence bounds provide another relevant formulation \cite{rink2025postselection}. Our evaluated procedure is a finite-family, rule-specific application using exact binomial lower bounds and Bonferroni control. We use this established risk-control construction to study its empirical deployment and reliability tradeoffs in bounded fraud-rule portfolios; the contribution is an application-focused evaluation, not a new general risk-control principle. Candidate rules use continuous score-like features and binary reason-code-like indicators. Score degradation is exogenous; it is not an adaptive attack and does not test evasion resistance.

This paper contributes a controlled empirical characterization of how selection-aware certification changes deployment and floor attainment as certification sample size, candidate pool, client floor, and distribution shift vary. We compare the certificate with point-estimate selection and a fixed safety-margin baseline on matched candidate pools, then examine candidate coverage in a chronological case study using the public synthetic BAF Base dataset. The study evaluates the certification policy rather than reproducing the full SpectralRules/PORS pipeline or an operational system; conclusions apply to the specified generators, sample sizes, candidate pools, and shift mechanisms.

\section{Related Work}
Wen and Lou's two-stage fraud-rule mining study is a close precedent. SpectralRules generates a compact candidate pool, and PORS searches Pareto-optimal OR rule subsets in precision--recall space. Their work establishes candidate-pool design, OR portfolios, and precision--recall-constrained selection as prior art \cite{wen2024pors}. The present study does not reproduce SpectralRules or PORS and does not compare its selector with PORS. Instead, it examines how a finite-family certification step changes deployment and floor attainment relative to point-estimate and fixed-margin selection on matched candidate pools.

Fraud-specific rule induction has also studied extracting compact rule sets from decision trees under low false-positive or alert-rate targets. RIFF extracts positive-class tree-path rules, then greedily selects rules by precision on uncovered examples to maximize recall under a false-positive rate (FPR) or alert-rate limit. Its experiments include the synthetic BAF account-fraud suite and time-based evaluation for BAF and industry data \cite{martins2024riff,jesus2022baf}. Vorobyev and Krivitskaya generate rules with tree-based models using components of existing expert rules as features, targeting reduced false positives while maintaining fraud-capture performance in a bank setting \cite{vorobyev2022ruleinduction}. MetaRule combines neural and tree-derived candidates and performs greedy rule-set learning to maximize fraud recall under a predefined criterion \cite{yu2022metarule}. These works establish substantial prior art in automatic fraud-rule generation, rule-derived inputs, portfolio selection, and constrained objectives. The present study compares readable conjunctions with CART positive-leaf candidates and evaluates selection-aware certification over their finite OR-portfolio families. It does not reproduce RIFF's rule-selection procedure or evaluate MetaRule or SpectralRules Stage 1.

Two earlier systems also address expert-facing fraud-rule workflows. RUDOLF proposes candidate adaptations to existing rules and iteratively incorporates expert feedback as transaction patterns evolve \cite{milo2016rudolf}. GOLDRUSH adapts rules across financial-institution contexts using semantic mappings and an integer-programming cost-benefit objective; its formulation allows precision, recall, and captured-legitimate limits \cite{jarovsky2018goldrush}. RUDOLF centers on iterative refinement, and GOLDRUSH on cross-context transfer. This prior work establishes client-driven workflows, candidate recommendations, and constrained rule adaptation. The distinct contribution here is a controlled empirical evaluation of how selection-aware certification affects deployment and floor attainment for a fixed family of bounded OR portfolios; the paper does not claim a new workflow concept or objective.

ARMS evaluates rule contributions and searches active configurations using a user-defined loss function, priorities, and operational actions; its paper reports synthetic experiments and two real-client applications \cite{aparicio2020arms}. ARMS manages an existing expert-authored rule system, whereas this study evaluates certification of a finite family of automatically generated OR portfolios. We do not make a head-to-head performance claim against ARMS.

Learn-then-Test (LTT) is a close precedent: it frames risk control over a finite family as multiple hypothesis testing and uses calibration data to identify procedures with controlled risk \cite{angelopoulos2025learntest}. Risk-controlling prediction sets and post-selection confidence bounds provide related finite-sample risk-control formulations \cite{bates2021riskcontrol,rink2025postselection}. Our benchmark uses one-sided exact binomial lower bounds with Bonferroni control for precision or recall of bounded OR portfolios. We do not compare against the complete LTT framework or establish a more general guarantee. The empirical question is how this conservative implementation trades deployment frequency against floor violations, particularly when the target population shifts away from the certification sample.

Reusable-holdout methods address validity under repeated adaptive queries, while weighted conformal prediction studies prediction-set coverage under covariate shift given appropriate shift weights \cite{dwork2015holdout,tibshirani2019covariateshift}. These are related but distinct settings: this benchmark evaluates one fixed finite portfolio family with a single certification step, and its binomial certificate does not correct for shift. We do not implement either approach.

Security work has examined evasion against banking fraud detectors under different attacker-knowledge assumptions and fraud-domain feature constraints, using raw transaction actions and aggregated behavioral features \cite{carminati2020evasion}. Other work studies black-box attacks on transaction sequences and tabular fraud data \cite{fursov2021adversarial,cartella2021adversarial}. These studies make clear that perturbations in a detector's feature space need not correspond to feasible raw transactions. The present generator contains already-aggregated synthetic score-like values and reason-code-like indicators, but no raw transaction variables or mapping from attacker actions to those signals. Directly editing these generated features would therefore grant unjustified control to a hypothetical attacker. We do not model evasion or claim adversarial robustness; the score degradation studied here is exogenous distribution shift. An evasion extension would require a defensible raw-action-to-feature map, linked-feature constraints, manipulation costs, and an explicit attacker goal and knowledge model.

Bayesian Rule Sets learn disjunctive normal form classifiers with user-controlled preferences for model size and shape \cite{wang2017brs}. CORELS studies certifiably optimal rule lists, a distinct representation from unordered ORs of AND-rules \cite{angelino2017corels}. These learned representations are not comparative baselines in the present experiments.

Synthetic transaction simulation can support research when sensitive transaction data cannot be shared. The public Bank Account Fraud (BAF) suite was generated from an anonymized account-opening fraud dataset and includes temporal dynamics and class imbalance \cite{jesus2022baf,kagglebaf2022}. We use its Base variant in one temporal case study, but this remains synthetic evidence and does not establish performance on real transactions. RIFF has also evaluated on BAF, so the dataset itself is not a novel contribution \cite{martins2024riff}. PaySim describes an agent-based mobile-money simulator calibrated with transaction logs \cite{lopezrojas2016paysim}. Synthetic financial transactions have also been released for anti-money-laundering research \cite{altman2023synthetic}. The two generators in this paper are narrower: they were designed to expose selected rule-coverage and target-attainment behaviors, were not calibrated to transaction logs, and do not estimate performance in a real fraud population.

\section{Problem Formulation}
Let a labeled dataset be $D=\{(x_i,y_i)\}_{i=1}^{n}$, with $y_i\in\{0,1\}$ denoting the observed synthetic fraud label. A candidate rule $r_j$ is a conjunction of one or two feature conditions. For a selected rule set $S$, the OR strategy is
\begin{equation}
 h_S(x_i)=\bigvee_{r_j\in S}r_j(x_i),\qquad
 C(S)=\{i:h_S(x_i)=1\}=\bigcup_{r_j\in S} C_j,
\end{equation}
where $C_j$ is the set matched by rule $r_j$. Each transaction contributes at most once to confusion counts. Precision and recall are calculated on $C(S)$, not by averaging the standalone metrics of selected rules.

For a precision-floor/recall-maximizing client, the study policy chooses
\begin{equation}
 \max_{S\in\mathcal{S}} \operatorname{Recall}(S)
 \quad\text{subject to}\quad \operatorname{Precision}(S)\geq\tau_P.
 \label{eq:recall}
\end{equation}
For a recall-floor/precision-maximizing client, it chooses
\begin{equation}
 \max_{S\in\mathcal{S}} \operatorname{Precision}(S)
 \quad\text{subject to}\quad \operatorname{Recall}(S)\geq\tau_R.
 \label{eq:precision}
\end{equation}
Here $\mathcal{S}$ is the study's candidate strategy space, and $\tau_P$ and $\tau_R$ are prespecified target values. No alert-volume or false-positive cap is imposed in this study. Alert rate and false-positive rate are reported as secondary outcomes. A four-rule limit and a top-10 shortlist are computational design choices for the benchmark, not claimed production limits.

\section{Study Design}
\subsection{Synthetic generators}
We evaluate two author-designed generators that expose controlled candidate-rule overlap, class prevalence, label noise, and distribution shift. The first generator has complementary and redundant reason-code structures crossed with a $2^5$ factorial grid. The corrected follow-up uses 75 seeds per structure-condition cell, for 4,800 datasets. Each dataset has 3,000 training, 1,500 validation, and 3,000 held-out examples. Latent prevalence is 0.03 or 0.07; observed labels have 0 or 0.05 independent flip probability; score strength and reason sensitivity have two levels; and held-out score degradation is either absent or moderate. The two reason-code structures receive equal weight. The second generator is chronological: 1,600 datasets with five training months, one validation month, and two held-out months of 1,000 records each. Thus, the full and half certification samples contain 1,500 and 750 records in the first generator, and 1,000 and 500 records in the chronological generator. Neither generator is fitted to customer or production data.

\subsection{Candidate generation and selection}
Before validation, each training sample is deterministically divided into candidate-fit (80\%) and ranking (20\%) partitions. The readable pool contains positive single-feature conditions and two-condition conjunctions across distinct features, with numeric thresholds at training quantiles and minimum support of 20. The CART pool is fit on the candidate-fit partition; its leaf budget is selected by average precision on the ranking partition, and positive-leaf paths form the candidates. For each pool and objective, candidates are ranked on the ranking partition and the top 10 are frozen. All exact searches consider nonempty OR portfolios of at most four rules. Every portfolio metric is calculated on the unique union of matched examples.

We compare five policies. \emph{Training point estimate} selects the best portfolio meeting the client floor on ranking data. \emph{Five-point margin} uses the same ranking data and shortlist but requires the observed constrained metric to exceed its floor by 0.05. \emph{Full-validation point estimate} ranks and selects on the complete validation set without certification, representing a data-rich conventional baseline. \emph{Half-validation certification} uses the training-ranked shortlist and reserves a random half of validation for certification. \emph{Full-validation certification} uses the same frozen shortlist and reserves all validation rows for certification. The latter two policies are directly comparable in candidate family and ranking data; their certification sample sizes differ.

For certification, precision uses true positives as successes and alerts as trials; recall uses true positives as successes and actual fraud labels as trials. We calculate a one-sided 95\% Clopper--Pearson lower bound for every portfolio and apply Bonferroni correction across $M=\sum_{k=1}^{4}{10\choose k}=385$ possible portfolios. A certified portfolio must have a lower bound at least equal to the specified client floor. The certificate relies on independent, exchangeable certification examples and a candidate family frozen before certification. It does not guarantee performance under a shifted target distribution.

\subsection{Client objectives and outcomes}
The precision-floor/recall-maximizing objective requires precision at least $\tau_P$ and maximizes recall; the recall-floor/precision-maximizing objective requires recall at least $\tau_R$ and maximizes precision. We use $\tau_P\in\{0.60,0.75\}$ and $\tau_R\in\{0.30,0.50\}$ as design settings, not operational recommendations. We separately report (i) deployment rate, (ii) floor-violation rate among deployed portfolios, and (iii) overall floor attainment, where nondeployment is shown separately rather than treated as a deployed failure. Paired intervals resample complete factorial cells, preserving the design's heterogeneous conditions; effect sizes are reported directly without significance tests over the simulated replications.

\subsection{Validation-size follow-up}
To measure the effect of validation size, a separate prespecified sweep uses four stable balanced-generator settings (two reason-code structures by prevalence 0.03/0.07), 100 paired replicates per setting, and validation sizes of 1,500, 15,000, and 150,000. It uses the same training-derived candidate pools and top-10 shortlist, and compares full-validation point selection with simultaneous certification. Each replicate is evaluated on an independent 100,000-row held-out sample from the same stable mechanism. The sweep is restricted to these four synthetic settings; its results are descriptive rather than an estimate of deployment-population performance.

\subsection{Public synthetic case study}
The BAF Base analysis uses months 0--5 for development, month 6 for validation, and month 7 for held-out evaluation. Following the dataset's low fraud prevalence, the objective is maximize recall subject to a 5\% false-positive-rate cap. We split development data 80/20 for candidate fitting and ranking, use CART positive-leaf paths, and compare training point selection, a 1-percentage-point FPR margin, point selection on all of month 6, and full-month certification with an exact upper bound on FPR. This is a single public synthetic dataset, not an independent replication. The earlier precision-floor analysis was uninformative at this prevalence and is not used as efficacy evidence. BAF rows are not redistributed; the exact source release and checksum are recorded in the reproduction artifacts.

\begin{figure}[t]
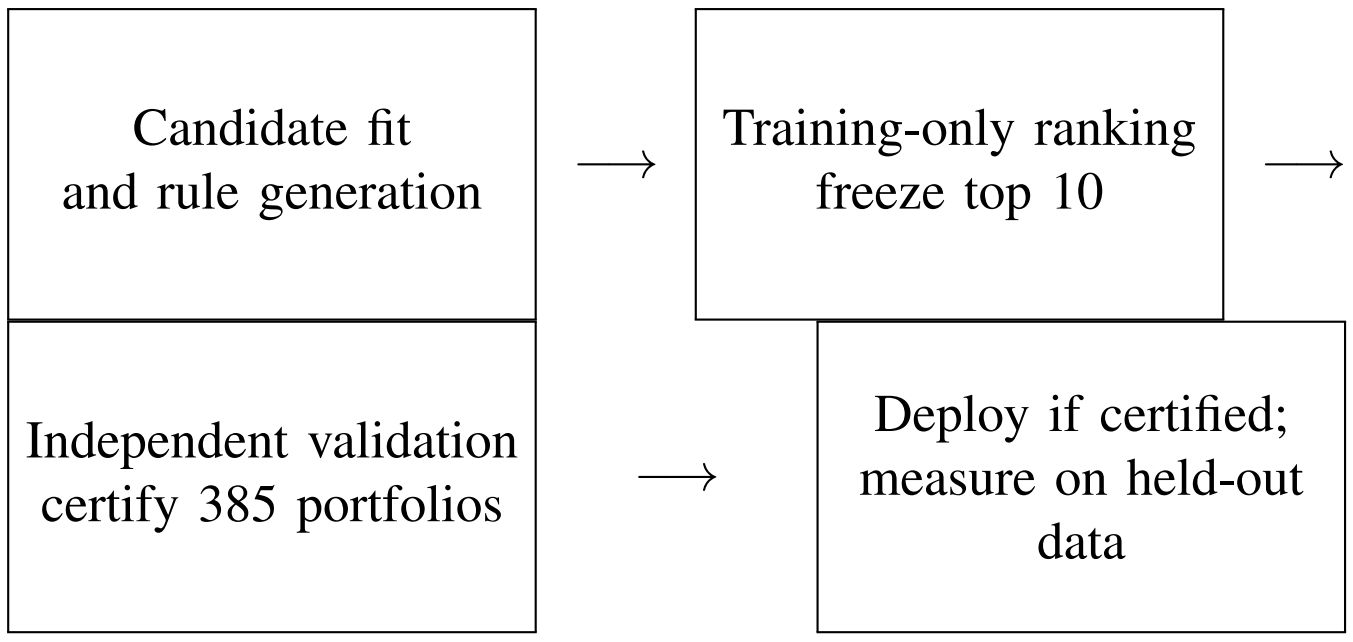

\centering
\fbox{\parbox[c][0.72in][c]{0.18\textwidth}{\centering Candidate fit\\and rule generation}}
\hfill $\longrightarrow$ \hfill
\fbox{\parbox[c][0.72in][c]{0.18\textwidth}{\centering Training-only ranking\\freeze top 10}}
\hfill $\longrightarrow$ \hfill
\fbox{\parbox[c][0.72in][c]{0.18\textwidth}{\centering Independent validation\\certify 385 portfolios}}
\hfill $\longrightarrow$ \hfill
\fbox{\parbox[c][0.72in][c]{0.18\textwidth}{\centering Deploy if certified;\\measure on held-out data}}
\caption{Evaluation workflow. Candidate generation and ranking are completed before the certification sample is used. Held-out data are used only for final evaluation.}
\label{fig:workflow}
\end{figure}

\section{Results}
\subsection{Deployment and floor reliability}
Table~\ref{tab:v12_cell_results} reports deployment rate and floor attainment conditional on deployment for every generator, candidate pool, and client floor. The full-validation certificate increased deployment relative to half-validation certification in all 16 cells. The average increase was 17.1 percentage points; average all-case floor attainment increased 12.2 points. The conditional violation rate, however, was 3.1 points higher on average, with substantial variation across cells. More certification data made certification less selective, but did not uniformly reduce violations on the shifted held-out samples.

Relative to point-estimate selection on all validation data, full-validation certification reduced deployment by 31.7 points on average and reduced the conditional floor-violation rate by 29.1 points; average all-case attainment increased by 3.0 points. Relative to the five-point safety-margin selector, full-validation certification reduced deployment by 29.4 points and the conditional violation rate by 22.2 points, while average all-case attainment decreased by 2.4 points. In the condition-cluster bootstrap, the interval for the conditional violation-rate difference was entirely below zero in 14 of 16 cells for each point-estimate baseline; two cells in each comparison had intervals overlapping zero. These equal-cell averages summarize this factorial design, not an operational population.

The results do not establish that certification is universally preferable. Full-validation certification substantially reduces violations among deployed portfolios, but deployment and all-case attainment can fall. The fixed 0.05 margin also reduces the violation rate while generally retaining more deployments. The choice depends on the cost of a missed floor versus abstention, and the study does not assign client-specific costs.

\subsection{Validation-size sensitivity}
Table~\ref{tab:validation_size} summarizes the separate no-shift sweep. Across its 32 equally weighted DGP-by-pool-by-floor cells, certification deployment increased from 61.3\% at 1,500 validation rows to 90.2\% at 15,000 and 92.5\% at 150,000. No held-out floor violations were observed among the deployed certificates on the independent 100,000-row samples (0/1,962; 0/2,887; and 0/2,961, respectively). Point selection on the full validation sample deployed in 97.5\%, 96.7\%, and 95.9\% of cases, while its equal-cell mean conditional violation rate declined from 20.6\% to 9.9\% and 4.5\%. Cell-level ranges remain wide, especially for point selection, so these averages are descriptive and do not imply general coverage or a universal sample-size threshold.

\begin{table}[t]
\centering
\scriptsize
\caption{Validation-size sweep across 32 equally weighted DGP-by-pool-by-floor cells, with 100 paired replicates per cell. Deployment and all-case attainment are equal-cell means; ranges show minima and maxima across cells. Conditional violation is the equal-cell mean among cells with at least one deployment; no certification violations were observed in this sweep.}
\label{tab:validation_size}
\resizebox{\columnwidth}{!}{%
\begin{tabular}{llccc}
\toprule
Validation $n$ & Method & Deploy (mean [range]) & Violation (mean [range]) & All-case \\
\midrule
1,500 & Point estimate & 97.5 [72, 100] & 20.6 [0, 90.3] & 78.3 \\
1,500 & Certification & 61.3 [0, 100] & 0.0 [0, 0] & 61.3 \\
15,000 & Point estimate & 96.7 [55, 100] & 9.9 [0, 60.9] & 88.2 \\
15,000 & Certification & 90.2 [0, 100] & 0.0 [0, 0] & 90.2 \\
150,000 & Point estimate & 95.9 [39, 100] & 4.5 [0, 37.9] & 92.4 \\
150,000 & Certification & 92.5 [0, 100] & 0.0 [0, 0] & 92.5 \\
\bottomrule
\end{tabular}%
}
\end{table}

\subsection{Exploratory population-reference diagnostic}
As a post hoc implementation check, we sampled up to five deployed full-validation certificates per stable balanced-generator condition, candidate pool, and client floor, yielding 999 portfolios across 32 DGP conditions. We independently parsed the frozen rule expressions and evaluated each unique OR union on 250,000 fresh synthetic examples per condition. For all 999 portfolios, the certification lower bound was below the two-sided 95\% Monte Carlo reference interval in 998 cases and overlapped it in one; none exceeded the interval. This diagnostic finds no obvious implementation discrepancy in this sample. Because it was conducted after the benchmark, reuses the author-designed generator helper, and samples only deployed certificates, it is not an independent or confirmatory estimate of familywise coverage.

\begin{table*}[t]
\centering
\scriptsize
\setlength{\tabcolsep}{4pt}
\caption{Results by generator, candidate pool, and client floor. Each cell reports deployment rate / held-out floor attainment among deployed portfolios (percent). Nondeployment is reported separately and is not counted as a deployed violation. ``Point full'' uses the complete validation sample for ranking and selection; ``Margin'' adds 0.05 to the client floor; ``Cert 1/2'' and ``Cert full'' use half or all of validation for certification.}
\label{tab:v12_cell_results}
\resizebox{\textwidth}{!}{%
\begin{tabular}{lllccccc}
\toprule
Generator & Pool & Floor & Point train & Margin & Point full & Cert 1/2 & Cert full \\
\midrule
Balanced & Shallow & $P\geq0.60$ & 99.7/38.9 & 99.1/51.6 & 99.9/39.2 & 33.8/98.5 & 50.9/96.2 \\
Balanced & Shallow & $P\geq0.75$ & 96.1/25.8 & 91.5/35.2 & 94.2/29.8 & 6.2/95.3 & 24.2/95.9 \\
Balanced & Shallow & $R\geq0.30$ & 100.0/52.7 & 100.0/65.1 & 100.0/52.6 & 96.5/94.0 & 99.3/89.2 \\
Balanced & Shallow & $R\geq0.50$ & 100.0/39.4 & 100.0/53.9 & 100.0/44.7 & 96.3/92.0 & 99.7/86.9 \\
Balanced & CART & $P\geq0.60$ & 99.9/91.0 & 99.7/91.7 & 100.0/90.8 & 49.5/98.9 & 75.4/97.8 \\
Balanced & CART & $P\geq0.75$ & 98.6/74.5 & 97.2/77.3 & 99.4/73.9 & 15.4/97.6 & 41.0/94.9 \\
Balanced & CART & $R\geq0.30$ & 84.7/45.5 & 77.5/48.7 & 85.2/45.3 & 50.7/55.8 & 62.0/52.9 \\
Balanced & CART & $R\geq0.50$ & 57.5/41.0 & 50.8/46.3 & 57.0/41.9 & 23.2/51.3 & 34.7/50.1 \\
Chronological & Shallow & $P\geq0.60$ & 100.0/32.4 & 100.0/40.3 & 100.0/33.1 & 28.6/84.0 & 49.9/69.6 \\
Chronological & Shallow & $P\geq0.75$ & 99.2/23.4 & 97.5/31.7 & 99.2/24.0 & 3.1/70.0 & 23.2/73.3 \\
Chronological & Shallow & $R\geq0.30$ & 100.0/49.9 & 100.0/66.6 & 100.0/47.9 & 92.2/97.8 & 98.6/94.6 \\
Chronological & Shallow & $R\geq0.50$ & 100.0/47.4 & 100.0/64.6 & 100.0/47.2 & 83.1/98.1 & 99.6/95.7 \\
Chronological & CART & $P\geq0.60$ & 100.0/81.8 & 100.0/81.8 & 100.0/81.6 & 43.6/96.6 & 76.9/90.3 \\
Chronological & CART & $P\geq0.75$ & 99.9/62.2 & 99.9/62.8 & 100.0/62.9 & 7.2/79.1 & 37.7/78.8 \\
Chronological & CART & $R\geq0.30$ & 90.1/48.8 & 81.1/53.2 & 90.6/49.4 & 48.2/62.9 & 63.9/58.1 \\
Chronological & CART & $R\geq0.50$ & 59.7/44.4 & 53.2/48.5 & 60.1/44.6 & 26.7/51.1 & 40.7/49.8 \\
\bottomrule
\end{tabular}%
}
\end{table*}

\begin{table}[t]
\centering
\small
\caption{Equal-cell average differences for full-validation certification versus each comparator across 16 generator-by-pool-by-floor cells (percentage points). Conditional violation is the failure rate among deployed portfolios.}
\label{tab:v12_tradeoff}
\resizebox{\columnwidth}{!}{%
\begin{tabular}{lrrrc}
\toprule
Comparator & $\Delta$ deploy & $\Delta$ all-case & $\Delta$ violation & Cell intervals below 0 \\
\midrule
Point, full validation & $-31.7$ & $+3.0$ & $-29.1$ & 14/16 \\
Point, +0.05 margin & $-29.4$ & $-2.4$ & $-22.2$ & 14/16 \\
Certification, half validation & $+17.1$ & $+12.2$ & $+3.1$ & 0/16 \\
\bottomrule
\end{tabular}%
}
\end{table}

\begin{figure*}[t]
\centering
\setlength{\unitlength}{0.022in}
\begin{picture}(292,93)
\put(16,15){\line(1,0){100}}\put(16,15){\line(0,1){60}}
\put(66,7){\makebox(0,0){\tiny Deployment rate (\%)}}
\put(2,45){\makebox(0,0){\rotatebox{90}{\tiny Floor violation among deployed (\%)}}}
\put(66,83){\makebox(0,0){\small Balanced generator}}
\put(16,12){\makebox(0,0){\tiny 0}}\put(66,12){\makebox(0,0){\tiny 50}}\put(116,12){\makebox(0,0){\tiny 100}}
\put(12,15){\makebox(0,0){\tiny 0}}\put(12,45){\makebox(0,0){\tiny 50}}\put(12,75){\makebox(0,0){\tiny 100}}
\put(108,44){\color{black}\circle*{3}}\put(111,47){\tiny Train point}
\put(105,39){\color{cyan}\circle*{3}}\put(109,37){\tiny +0.05 margin}
\put(108,43){\color{blue}\circle*{3}}\put(99,49){\tiny Full-val point}
\put(63,24){\color{orange}\circle*{3}}\put(67,23){\tiny Half cert}
\put(77,25){\color{red}\circle*{3}}\put(81,26){\tiny Full cert}
\put(166,15){\line(1,0){100}}\put(166,15){\line(0,1){60}}
\put(216,7){\makebox(0,0){\tiny Deployment rate (\%)}}
\put(166,12){\makebox(0,0){\tiny 0}}\put(216,12){\makebox(0,0){\tiny 50}}\put(266,12){\makebox(0,0){\tiny 100}}
\put(162,15){\makebox(0,0){\tiny 0}}\put(162,45){\makebox(0,0){\tiny 50}}\put(162,75){\makebox(0,0){\tiny 100}}
\put(216,83){\makebox(0,0){\small Chronological generator}}
\put(260,46){\color{black}\circle*{3}}\put(263,48){\tiny Train point}
\put(258,41){\color{cyan}\circle*{3}}\put(253,38){\tiny +0.05 margin}
\put(260,46){\color{blue}\circle*{3}}\put(249,52){\tiny Full-val point}
\put(208,27){\color{orange}\circle*{3}}\put(211,28){\tiny Half cert}
\put(227,29){\color{red}\circle*{3}}\put(230,30){\tiny Full cert}
\end{picture}
\caption{Deployment--violation tradeoff. Each point is the equal-cell average over two candidate pools and four client floors within the named generator. Certification increases reliability among deployed cases but reduces deployment; using the full validation sample partly recovers deployment.}
\label{fig:tradeoff}
\end{figure*}
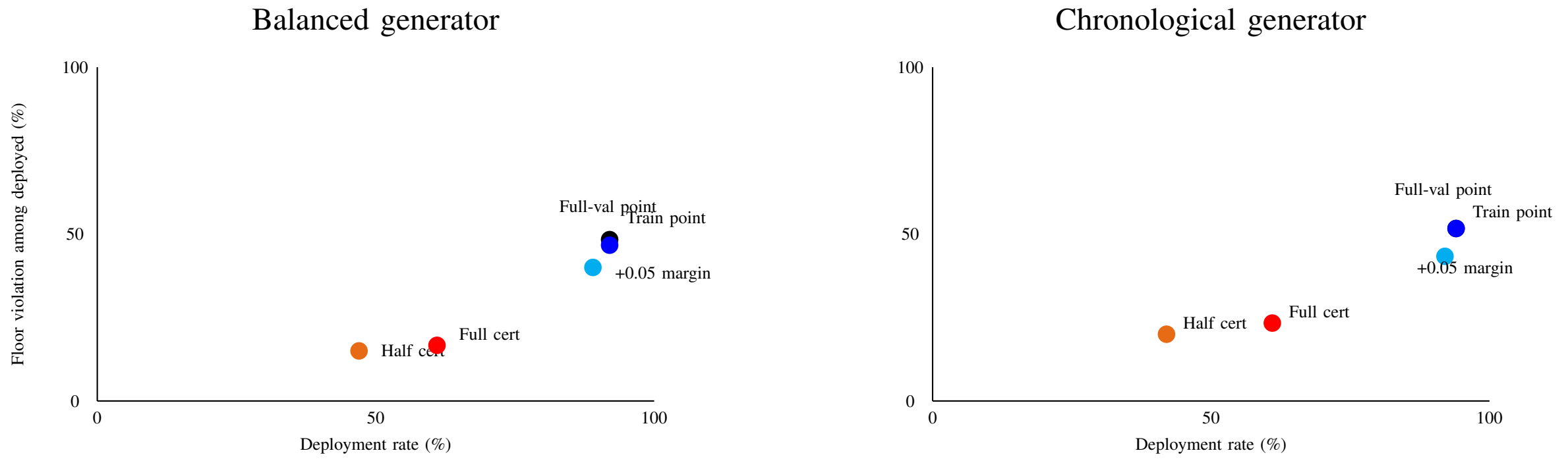

\subsection{BAF Base at a 5\% false-positive-rate cap}
The BAF candidate pool produced only five positive CART paths. All four policies selected a portfolio, and all had held-out FPR below 0.02\%, far under the 5\% cap. Test recall ranged from 0.77\% for full-validation point selection to 0.91\% for the training point and certified selectors; the certified selector's simultaneous upper bound on validation FPR was 0.037\%. The FPR constraint was therefore nonbinding for this small candidate pool, and the very low recall indicates limited candidate coverage. Because this is one fixed temporal test set, these figures are descriptive and do not provide independent external validation.

\begin{table}[t]
\centering
\small
\caption{BAF Base, months 0--5/6/7. One fixed temporal split; no population-level intervals.}
\label{tab:baf_fpr5}
\resizebox{\columnwidth}{!}{%
\begin{tabular}{lcccc}
\toprule
Policy & Deploy & Test recall & Test FPR & Test precision \\
\midrule
Training point & Yes & 0.91\% & 0.0147\% & 48.15\% \\
1-point margin & Yes & 0.91\% & 0.0147\% & 48.15\% \\
Full-validation point & Yes & 0.77\% & 0.0157\% & 42.31\% \\
Full-validation certificate & Yes & 0.91\% & 0.0147\% & 48.15\% \\
\bottomrule
\end{tabular}%
}
\end{table}

\section{Reproducibility}
The primary follow-up contains 6,400 datasets and 256,000 seed-level method rows. Its frozen design protocol, source files, environment lock, seed-level outcomes, summaries, condition-cluster bootstrap contrasts, and structural audit are included in the artifact bundle. A post hoc analysis change replaced planned cell-mean t intervals with condition-cluster bootstrap intervals to account for the factorial structure; point estimates are unchanged, and inferential claims are kept descriptive. The audit recomputes the exact bounds and verifies row structure; it shares generator and metric helpers with the benchmark. A separately implemented rule parser supports the exploratory reference audit. The validation-size sweep includes 400 datasets and 19,200 method rows. BAF aggregate results, source checksum, environment record, and bound audit are included; row-level rule expressions and the raw CSV are not redistributed.

\section{Limitations and Responsible Use}
The two synthetic generators are author-designed and are not calibrated to representative transaction logs. The chronological generator and the shared candidate/selector code were developed by the same research team. The balanced follow-up quantifies behavior across two distinct reason-code structures, but does not cover independently developed generators or real fraud populations. The condition-cluster bootstrap summarizes uncertainty within this factorial design; transfer across independently developed generators remains unmeasured. A post hoc Monte Carlo diagnostic checks sampled certificates against reference metrics from the same synthetic mechanisms. It is an implementation diagnostic, not an independent confirmatory coverage study; the certificate's stated guarantee continues to rely on the finite-sample binomial assumptions described above.

The certificate applies to the fixed finite candidate family when certification examples are independent and exchangeable with the population for which the floor is asserted. It is not distribution-free under arbitrary shift. The deliberately shifted held-out samples show that a valid certificate on validation data can coexist with later floor violations. The five-point baseline is one prespecified margin and does not exhaust practical alternatives; we do not implement the complete Learn-then-Test framework, fixed-sequence testing, or importance-weighted certification.

The BAF Base case study is one temporal split of synthetic data. Its CART pool contains only five candidate rules, and held-out recall remains below 1\% while the 5\% FPR cap is far from binding. This makes the case study a candidate-coverage diagnostic for this split rather than a measure of practical fraud-capture utility. No row-level BAF data are redistributed. The study models no adaptive attacker, delayed labels, selective feedback, or production detector and supports no claim of evasion robustness, security efficacy, or operational performance.

\section{Broader Impact}
Fraud detection can reduce financial losses and investigation burden, but false positives can block legitimate applicants and may affect groups unevenly. This study uses synthetic data, does not measure subgroup impacts, and does not evaluate a deployed decision system. Its certificates depend on exchangeability with the validation population and may fail under distribution shift. Any operational use would require human review, subgroup and customer-impact monitoring, drift checks, and an appeal path; these synthetic findings alone are not a basis for deployment.

\section{Conclusion}
Across the controlled synthetic settings, full-validation certification trades lower floor-violation rates among deployed portfolios for lower deployment frequency than point-estimate and fixed-margin selection. Using the full validation sample rather than half of it raises deployment frequency, but its effect on held-out violations varies under distribution shift. The single BAF case study has low recall and a nonbinding FPR cap, making it informative about candidate coverage in this split but not about practical fraud-capture value. The main finding is a measured reliability--deployment tradeoff within the tested generators and candidate pools; performance on real transaction populations and broader method comparisons remain open evaluation questions.

\section*{Acknowledgment}
OpenAI Codex was used as follows. Code: it assisted in writing and debugging simulation code and analysis scripts; separate audit scripts recomputed certification bounds and portfolio metrics, while some generator, rule-application, and metric helpers were shared with the benchmark. Literature: it suggested candidate references, which the authors checked against the original sources. Text: it produced draft text for the Abstract, which the authors revised and rewrote, and assisted with language editing of selected passages. Figures: Codex-assisted LaTeX code was used for the workflow schematic (Figure~1) and trade-off visualization (Figure~2), whose plotted values came from study results. The authors designed the research, reviewed the results, and are responsible for the content. No customer transaction data were used; this study used author-generated synthetic data and the public BAF dataset. The authors declare no conflicts of interest. This work received no external funding.

\section*{Data Availability}
The benchmark source code, frozen protocols, environment records, synthetic results, analysis scripts, and audit reports are provided in the supplementary artifact package accompanying this manuscript. The BAF Base CSV and row-level rule expressions are not redistributed; the dataset is available from Kaggle Version 2 \cite{kagglebaf2022}.

\begin{IEEEbiographynophoto}{Rishabh Pande}
Rishabh Pande is a Senior Director of Data Science at Socure, an identity verification company. His work focuses on machine learning for identity verification and fraud detection, including neural networks, transformer models, graph methods, and AI agents. He works with client data to evaluate the value of data science products. His research interests include applied machine learning, fraud analytics, and responsible AI.
\end{IEEEbiographynophoto}

\begin{IEEEbiographynophoto}{Yu Zeng}
Yu Zeng is a Lead Data Scientist at Socure and holds a Ph.D. His work includes machine learning for identity verification and fraud detection. His research interests include neural networks, transformer models, graph learning, and AI agents.
\end{IEEEbiographynophoto}


\begin{thebibliography}{99}

\bibitem{openai_codex}
OpenAI, ``Codex,'' 2026. [Online]. Available: \url{https://openai.com/codex}. Accessed: Oct. 20, 2026.

\bibitem{cawley2010selection}
G. C. Cawley and N. L. C. Talbot, ``On over-fitting in model selection and subsequent selection bias in performance evaluation,'' \emph{J. Mach. Learn. Res.}, vol. 11, no. 70, pp. 2079--2107, 2010.

\bibitem{varma2006bias}
S. Varma and R. Simon, ``Bias in error estimation when using cross-validation for model selection,'' \emph{BMC Bioinformatics}, vol. 7, Art. no. 91, 2006, doi: 10.1186/1471-2105-7-91.

\bibitem{angelopoulos2025learntest}
A. N. Angelopoulos, S. Bates, E. J. Cand\`es, M. I. Jordan, and L. Lei, ``Learn then Test: Calibrating predictive algorithms to achieve risk control,'' \emph{Ann. Appl. Stat.}, vol. 19, no. 2, pp. 1641--1662, 2025, doi: 10.1214/24-AOAS1998.

\bibitem{rink2025postselection}
P. Rink and W. Brannath, ``Post-selection confidence bounds for prediction performance,'' \emph{Machine Learning}, vol. 114, Art. no. 82, 2025, doi: 10.1007/s10994-024-06632-w.

\bibitem{wen2024pors}
C. Wen and Y. Lou, ``On Finding Bi-objective Pareto-optimal Fraud Prevention Rule Sets for Fintech Applications,'' in \emph{Proc. 30th ACM SIGKDD Conf. Knowledge Discovery and Data Mining}, 2024, pp. 5959--5968, doi: 10.1145/3637528.3671521.

\bibitem{martins2024riff}
J. L. Martins, J. Bravo, A. S. Gomes, C. Soares, and P. Bizarro, ``RIFF: Inducing Rules for Fraud Detection from Decision Trees,'' in \emph{Proc. RuleML+RR}, 2024. [Online]. Available: \url{https://arxiv.org/abs/2408.12989}

\bibitem{jesus2022baf}
S. Jesus, J. Pombal, D. Alves, A. Cruz, P. Saleiro, R. P. Ribeiro, J. Gama, and P. Bizarro, ``Turning the Tables: Biased, Imbalanced, Dynamic Tabular Datasets for ML Evaluation,'' in \emph{Advances in Neural Information Processing Systems}, vol. 35, 2022. [Online]. Available: \url{https://arxiv.org/abs/2211.13358}

\bibitem{vorobyev2022ruleinduction}
I. A. Vorobyev and A. Krivitskaya, ``Reducing false positives in bank anti-fraud systems based on rule induction in distributed tree-based models,'' \emph{Computers \& Security}, vol. 120, Art. no. 102786, 2022, doi: 10.1016/j.cose.2022.102786.

\bibitem{yu2022metarule}
L. Yu et al., ``MetaRule: A Meta-path Guided Ensemble Rule Set Learning for Explainable Fraud Detection,'' in \emph{Proc. ACM CIKM}, 2022, pp. 4650--4654, doi: 10.1145/3511808.3557641.

\bibitem{milo2016rudolf}
T. Milo, S. Novgorodov, and W.-C. Tan, ``RUDOLF: Interactive Rule Refinement System for Fraud Detection,'' \emph{Proc. VLDB Endowment}, vol. 9, no. 13, pp. 1465--1468, 2016, doi: 10.14778/3007263.3007270.

\bibitem{jarovsky2018goldrush}
A. Jarovsky, T. Milo, S. Novgorodov, and W.-C. Tan, ``Rule Sharing for Fraud Detection via Adaptation,'' in \emph{Proc. IEEE 34th Int. Conf. Data Engineering (ICDE)}, 2018, pp. 125--136, doi: 10.1109/ICDE.2018.00021.

\bibitem{aparicio2020arms}
D. Apar\'icio, R. Barata, J. Bravo, J. T. Ascens\~ao, and P. Bizarro, ``ARMS: Automated rules management system for fraud detection,'' arXiv:2002.06075, 2020. [Online]. Available: \url{https://arxiv.org/abs/2002.06075}

\bibitem{bates2021riskcontrol}
S. Bates, A. Angelopoulos, L. Lei, J. Malik, and M. I. Jordan, ``Distribution-free, risk-controlling prediction sets,'' \emph{J. ACM}, vol. 68, no. 6, Art. no. 43, 2021, doi: 10.1145/3478535.

\bibitem{dwork2015holdout}
C. Dwork, V. Feldman, M. Hardt, T. Pitassi, O. Reingold, and A. Roth, ``The reusable holdout: Preserving validity in adaptive data analysis,'' \emph{Science}, vol. 349, no. 6248, pp. 636--638, 2015, doi: 10.1126/science.aaa9375.

\bibitem{tibshirani2019covariateshift}
R. J. Tibshirani, R. Foygel Barber, E. J. Cand\`es, and A. Ramdas, ``Conformal prediction under covariate shift,'' in \emph{Advances in Neural Information Processing Systems}, vol. 32, 2019.

\bibitem{carminati2020evasion}
M. Carminati, L. Santini, M. Polino, and S. Zanero, ``Evasion attacks against banking fraud detection systems,'' in \emph{Proc. 23rd Int. Symp. Research in Attacks, Intrusions and Defenses (RAID)}, 2020, pp. 285--300. [Online]. Available: \url{https://www.usenix.org/conference/raid2020/presentation/carminati}

\bibitem{fursov2021adversarial}
I. Fursov et al., ``Adversarial attacks on deep models for financial transaction records,'' in \emph{Proc. 27th ACM SIGKDD Conf. Knowledge Discovery and Data Mining}, 2021, pp. 2868--2878, doi: 10.1145/3447548.3467145.

\bibitem{cartella2021adversarial}
F. Cartella, O. Anunciacao, Y. Funabiki, D. Yamaguchi, T. Akishita, and O. Elshocht, ``Adversarial attacks for tabular data: Application to fraud detection and imbalanced data,'' in \emph{Proc. Workshop on Artificial Intelligence Safety (SafeAI)}, CEUR Workshop Proceedings, vol. 2808, 2021. [Online]. Available: \url{https://ceur-ws.org/Vol-2808/}

\bibitem{wang2017brs}
T. Wang, C. Rudin, F. Doshi-Velez, Y. Liu, E. Klampfl, and P. MacNeille, ``A Bayesian Framework for Learning Rule Sets for Interpretable Classification,'' \emph{J. Mach. Learn. Res.}, vol. 18, no. 70, pp. 1--37, 2017.

\bibitem{angelino2017corels}
E. Angelino, N. Larus-Stone, D. Alabi, M. Seltzer, and C. Rudin, ``Learning Certifiably Optimal Rule Lists,'' in \emph{Proc. ACM SIGKDD}, 2017, pp. 35--44.

\bibitem{kagglebaf2022}
S. Jesus, ``Bank Account Fraud Dataset Suite (NeurIPS 2022),'' Kaggle, Version 2, 2022. [Online]. Available: \url{https://www.kaggle.com/datasets/sgpjesus/bank-account-fraud-dataset-neurips-2022}

\bibitem{lopezrojas2016paysim}
E. A. Lopez-Rojas, A. Elmir, and S. Axelsson, ``PaySim: A financial mobile money simulator for fraud detection,'' in \emph{Proc. European Modeling and Simulation Symposium}, 2016, pp. 249--255. [Online]. Available: \url{https://www.msc-les.org/proceedings/emss/2016/EMSS2016_249.pdf}

\bibitem{altman2023synthetic}
E. Altman, J. Blanu\v{s}a, L. von Niederh\"ausern, B. Egressy, A. Anghel, and K. Atasu, ``Realistic Synthetic Financial Transactions for Anti-Money Laundering Models,'' arXiv:2306.16424, 2023. [Online]. Available: \url{https://arxiv.org/abs/2306.16424}
\end{thebibliography}
\end{document}